\documentstyle[preprint,eqsecnum,aps]{revtex}
\begin{document}
\title{Two-dimensional $XY$ spin/gauge glasses on periodic and quasiperiodic
       lattices}
\author{R.W. Reid, S.K. Bose and B. Mitrovi\'{c}}                     
\address{Physics Department, Brock University, St. Catharines, Ontario
         L2S 3A1 CANADA}
\date{\today}
\maketitle
\begin{abstract}
Via Monte Carlo studies of the frustrated XY or classical planar model we
demonstrate the possibility of a finite (nonzero) temperature spin/gauge glass
phase in two dimensions. Examples of both periodic and quasiperiodic
two dimensional lattices, where a high temperature paramagnetic phase changes
to a spin/gauge glass phase with the lowering of temperature, are presented.
The existence of the spin/gauge glass phase is substantiated by our study of the
temperature dependence of the Edwards-Anderson order parameter,
spin glass susceptibility, linear susceptibility and the specific heat. 
Finite size scaling analysis of spin glass susceptibility and order
parameter yields a nonzero critical temperature and exponents that are in close
agreement with those obtained by Bhatt and Young in their random ${\pm J}$
Ising model study on a square lattice. These results suggest that certain
periodic and  quasiperiodic two-dimensional arrays of superconducting grains
in suitably chosen transverse magnetic fields should behave as superconducting
glasses at low temperatures.
%
\end{abstract}
\pacs{75.10.Hk, 75.10.Nr, 64.70.Pf}
\section{Introduction}    
\label{sec:intro}
In a recent communication \cite{reid} in this journal, we reported
that the frustrated XY model (see Eq.(1)) on a two-dimensional (2D) Penrose
lattice \cite{strandburg} exhibits a low temperature spin/gauge glass phase
\cite{chowdhury,fradkin}.
In this work, we show that quasiperiodicity is not a necessary requirement
of the spin glass phase in this model.
This conclusion is based on our Monte Carlo (MC) study of the above model on
 the periodic honeycomb and bathroom tile lattices. When fully frustrated,
these lattices appear to undergo a paramagnetic to spin/gauge glass transition at a low, but
finite temperature, $T_f\,$. In addition, we find that the octagonal
quasiperiodic lattice \cite{soma} exhibits a   behavior  similar to that
of  the other three lattices, but with a somewhat higher $T_f\,$.

These results are unexpected and controversial in the light of the prevalent
notion that the critical dimension for a spin glass phase is greater than two
\cite{bhatt,jain,binder,morgenstern}. 
The validity of this notion, however, has been
put to doubt recently not only by our work \cite{reid} but also by the work of 
Lemke and Campbell \cite{lemke} in their study of the Ising model
with nearest neighbor random ${\pm \lambda J}$ and next nearest neighbor ferromagnetic
interactions on square lattices.
An important distinction between our study and that of Lemke and Campbell
\cite{lemke}
is that in the latter case frustration
is induced via (or accompanied by) disorder with the nearest neighbor
interactions varying randomly, whereas in our model frustration is nonrandom:
it is determined by the lattice structure and an applied magnetic field.
Thus our results provide definite examples of cases where frustration alone
(i.e., without disorder) is capable of inducing a spin glass phase. 
Our results clearly illustrate the role of lattice structure as a
relevant variable from the viewpoint of the universality class of the
transition. The frustrated XY model  has been studied widely on a variety of
periodic 2D  lattices using mean-field and renormalization group methods, and
Monte Carlo simulation  
\cite{teitel,shih,miya,lee-caf,yos,choi,dlee,berge,lee-kos,gran-kos,granato,jlee,ramirez,night}. 
The results, although sometimes ambiguous and often discrepant among various
authors, are certainly lattice-dependent. Two most commonly reported and
discussed transitions for the fully frustrated XY model are
the Ising type and the Kosterlitz-Thouless(KT) type \cite {koster,koster.74}.
We provide evidence that rule out the possibility of either of these
two types of transitions on the two periodic and the two quasiperiodic
lattices considered in this work.

The purpose of this work is two-fold: to show that both periodic and
quasiperiodic lattices can exhibit finite temperature spin glass transitions
in 2D, and  to provide some details of the study on the
Penrose lattice, which were left out of our previous communication \cite{reid}. With the example of the periodic  honeycomb and bathroom tile
lattices we dispel any misconception we may have inadvertently generated in
ref.1 that quasiperiodicity is a necessary condition for the spin glass phase
in the model. Both the periodic and the quasiperiodic lattices studied in
this work seem to undergo a paramagnetic to spin glass transition at
temperatures that are low but certainly nonzero.

We have studied the current model without the frustration term (the $A_{ij}$
term in Eq.(1)), i.e. the standard XY model,
on the quasiperiodic 2D Penrose lattice. We find that the transition is the
Kosterlitz-Thoules(KT)-type \cite{koster} with the exponents
identical to those obtained by Tobochnik and Chester \cite{toboch} in their
study on the square lattice, but the transition temperature
is somewhat higher.
Thus in the absence of frustration (or variation in the sign of the
interactions) quasiperiodicity is an irrelevant variable, as expected. 
The details of the standard XY model study on 2D and 3D Penrose lattices
will be published elsewhere.

The remainder of this paper is divided into sections as follows. In Sec.\
\ref{sec:model} we discuss our model and some features of the lattices
considered in this work. In Sec.\ \ref{sec:results} we present the results
of the MC simulations. In Sec.\ \ref{sec:comp} we compare our results with
those obtained using other lattices. We also suggest what characteristic of
our model might be responsible for the spin
glass phase. In Sec.\ \ref{sec:con} we present our comments and conclusions.

\section{The Model}
\label{sec:model}
We consider the Hamiltonian for the XY model describing the interaction between
2D spin vectors with orientations $\theta_i$ and $\theta_j$ situated at
lattice sites $i$ and $j$ via a nearest neighbor coupling parameter $J$:
\begin{equation}
H = -J\sum_{[ij]}\cos(\theta_{i}-\theta_{j}+A_{ij}),
\end{equation}
where the summation is restricted to nearest neighbor pairs   $[ij]$. 
The parameter $A_{ij}$ controls the frustration in the model. In the context
of an array of superconducting grains, $\theta_i$ is the phase of the
superconducting order parameter at the grain
$i$ and the above Hamiltonian can be seen as describing the resulting Josephson
junction of the grains ``minimally coupled''
to a transverse magnetic field with vector potential ${\bf A}$ with
\begin{equation}
A_{ij} = \frac{{2\pi}}{\Phi_0}\int^{\bf r_j}_{\bf r_i}{\bf A}\cdot d{\bf l}\:.
\end{equation}
$\Phi_0$ is the elementary flux quantum ${\frac{hc}{2e}}$ associated with the
Cooper pairs, and ${\bf r_i}$ denotes the lattice sites.
Here the magnetic field acts as the source of frustration: an $A_{ij}$ which is
an odd multiple of $\pi$ essentially renders the bond $[ij]$ negative. 
 
The directed sum of $A_{ij}$ about a plaquette in a 2D lattice can be written
as $2\pi f$, where $f$ is the flux through the plaquette in units of $\Phi_0$.
The 2D Penrose lattice \cite{strandburg} is composed of two (fat and thin)
rhombic unit cells (plaquettes). The ratio of the areas of the fat and the thin
rhombuses in the Penrose lattice is the Golden Mean ($\tau$), which is an
irrational number($\frac{1+\sqrt{5}}{2}$). Thus only one set
of plaquettes can be fully frustrated at a time with a suitable choice of the
magnetic field giving $f=1/2$. The flux $f$ through the individual plaquettes
in the other set will then be an irrational number.
The octagonal lattice \cite{soma} consists of two unit cells, one square and
the other a thin rhombus, with a ratio of $\sqrt{2}$ in their areas.
Similar to the Penrose lattice case, fully frustrating one of the plaquettes
results in an irrational flux through the other.

Fig.1 displays the lattices used in this work along with the reference
$x$ and $y$  directions. The linear dimensions of the clusters used in the
simulation along the $x$ and $y$ directions, $L_x$ and $L_y$, and the 
corresponding numbers of sites in the clusters are given in TABLE I.

In Figs.~1(a) and 1(b) we show  sections of Penrose (decagonal) and
octagonal quasilattices. Self-similarity, or equivalently, the
inflation-deflation property of these two quasilattices are characterized by 
two irrational numbers, the Golden Mean $(\tau=\frac{1+\sqrt{5}}{2})$ and
the Silver Mean $(\sigma=1+\sqrt{2})$, respectively, which also dictate their
decagonal and octagonal bond orientational symmetry. The two quasilattices
can be generated via projections of 5D and 4D simple
hypercubic (periodic) lattices on to the physical 2D plane.
They have similar ring structure, both containing only even order rings.
The average number of nearest neighbors for both quasilattices is four,
as in a square lattice. But unlike the square lattice, the two quasilattices 
are characterized by variations in near neighbor environments. For the
Penrose (decagonal) lattice the number of nearest neighbors varies between
three and seven, whereas for the octagonal lattice  the number varies between
three and eight. In order to reduce the surface effects in our finite cluster
MC calculations we have used periodic boundary conditions. Rational
approximants of the two quasilattices, which can be repeated periodically,
can be obtained from the rational approximations of the irrational numbers,
the Golden and the Silver Means.  We follow a systematic way to generate
these periodic approximants as given by Lan\c{c}on and Billard\cite{lancon}.

Figs. 1(c) and (d) display sections of the two periodic lattices, honeycomb
and bathroom tile, considered in the present work. Note that both 
are non-Bravais lattices with the same number of nearest neighbors. 
The smallest unit cells that provide a Bravais lattice description of these
lattices involve two sites for the honeycomb lattice and four for the
bathroom tile. The honeycomb lattice consists of only one type of plaquettes
(hexagonal), while the bathroom tile lattice has two types, square and
octagonal, with an irrational  ratio of $4(1+\sqrt{2})$ in the areas
of the two plaquettes (octagonal to square). Both structures contain only even
order rings. 

Finally, a word about the nomenclature used to describe the model given by
Eq. (1). If the $A_{ij}$ are restricted to the values $0$ and $\pi$ and randomly
assume these values with equal probability, then the model becomes
the Edwards-Anderson ${\pm J}$ XY
spin glass, with random ferromagnetic ($A_{ij}=0$) or antiferromagnetic
($A_{ij}=\pi$) coupling between adjacent spins.  If the $A_{ij}$
are independent random variables, assuming all values between $0$ and 
$2\pi$, the model is referred to as the gauge or vortex  glass model
\cite{reger} and
is believed to belong to a different universality class, presumably because
it lacks the "reflection" symmetry, $\theta_i \rightarrow -\theta_i \forall i$.
In the present case the $A_{ij}$ are lattice structure-dependent.  But they
assume many different values in the interval between $0$ and 
$2\pi$, depending on the lattice structure. Instead of dwelling on the
fine differences between a spin glass and a gauge glass, we will use the
term spin glass throughout the remainder of the paper, especially in  
describing the phase transition itself. The quantities we study show
temperature variations similar to those observed experimentally in the
so-called spin glasses, hence the choice.

\section{Results of MC simulation}
\label{sec:results}
For the lattices which contain two different plaquettes, we present results for
the case where the plaquettes with the  smaller area are  fully frustrated. 
For the 
Penrose lattice these are the thin rhombohedral  plaquettes. For the octagonal
and the bathroom tile lattices these are, respectively, the rhombohedral and
the square plaquettes. Results for the other case, where the plaquettes with
the larger area are fully frustrated, are qualitatively similar. 
All our results are obtained via MC simulation based on the Metropolis
algorithm \cite{metropolis}, using periodic boundary conditions.
We have cooled our systems in a quasi-static manner, starting from a high
temperature $(T$(in units of $J$)$>2.0)$ random configuration and then heated
the system in the same quasi-static fashion. Since we performed the simulation
in $n$ blocks, the heating and cooling data are obtained by averaging over
these blocks, with the error bars representing the standard deviation,
obtained by dividing the square root of the  sum of squares of the deviations
from the mean by $\sqrt{n-1}$, instead of  $\sqrt{n}$.
We then perform a `grand average' over the heating and cooling data.

By associating  a spin ${\vec S_i}$ = $(\cos\theta_i,\sin\theta_i)$ with every
lattice site $i$, we can define the quantity:
\begin{equation}
\frac{1}{N}\sum_{i}^N\langle \vec S_{i} \rangle \:, 
\end{equation}
where ${\langle \hskip .5pc\relax \rangle}$ denotes a
canonical ensemble average at a temperature $T$, as the magnetic moment per
site at temperature $T$. Note that this quantity does not represent the actual
magnetic moment of  a cluster of superconducting grains in a transverse
magnetic field. 
However, defined as above, magnetic moment per lattice site calculated for
all the lattices studied is found to be small ($<0.02$)
over the entire temperature range. The magnitude of the moment decreases
steadily with the size of the cluster, suggesting that the magnetization
is strictly zero in the thermodynamic limit. This is an indication that
the ground state (more appropriately, the lowest temperature (0.02 $J$)
state studied via our MC simulation) retains the continuous O(2) symmetry of 
the Hamiltonian, and
there is no spontaneous breakdown into the discrete Z(2) symmetry, 
as would be the case for an Ising transition.
However, the vanishing of the magnetic moment at all temperatures does not
rule out the possibility of a KT transition. Below we present further analysis
in an attempt to determine the existence and nature of the transition(s) for
the various lattices.

\subsection{Edwards-Anderson Order parameter}
Since the spins appear to be disordered at all temperatures, it is appropriate
to explore the possiblity of spin freezing over  macroscopic time scales.
To study the freezing of the spins at the lattice sites we calculate     
the Edwards-Anderson \cite{edwards} order parameter. 
In Fig.\ \ref{q.comp} we show this order parameter defined by:
\begin{equation}
q_{EA} = 1/N\sum_{i}^N\langle \vec S_{i} \rangle ^2 .
\end{equation}
In a completely frozen system $q_{EA}$ is unity, while for a completely
ergodic system it is zero.
This order parameter shows a monotonic decrease with increasing temperature,
clearly vanishing at temperatures beyond 0.5 for all lattices.
The results shown in Fig.\ \ref{q.comp} were obtained by averaging
over 5 blocks of 60,000 configurations, generated after equilibrium was
achieved.
In all cases the order parameter is seen to vanish not abruptly, but
continuously with a long tail. This is a consequence of the finite system size.
It is expected that the tail region will decrease with increasing system size
and eventually disappear in the thermodynamic limit. 
We find that the tail persists, even for our largest systems
(e.g. $\sim$ 11,000 site cluster for the Penrose lattice) and, consequently, the
transition temperature, at which $q_{EA}$ goes to zero, cannot be
appropriately determined from Fig.\ \ref{q.comp}. Thus, we use other quantities
to analyse the transition and provide estimates of the transition temperatures
$T_f$ for the four lattices. In Fig.\ \ref{q.comp}, in addition to the lattice
sizes, we have included our estimates of the transition
temperatures for the corresponding lattice, obtained from a finite size
scaling analysis of spin glass susceptibility (to be described later).

Although we cannot accurately determine a $T_f$ from this method, it is clear
that all the systems studied have a
low temperature spin/gauge phase with a nonzero order parameter $q_{EA}$,
changing into a phase with zero $q_{EA}$ as the temperature is raised. 

\subsection{Linear Susceptibility}
Linear susceptibilities  per spin for the Penrose and honeycomb lattices, calculated from the fluctuations in the
magnetization (net magnetic moment $|m|$ for a lattice of $N$ sites),
\begin{equation}
\chi = \frac{\langle m^2 \rangle - \langle |m| \rangle^2}{Nk_BT},
\end{equation}
are shown in Fig.\ \ref{x.comp} (a) and (b), respectively.
At low temperatures $(0.02 - 0.2)J$, the results are obtained by averaging 
over 5 blocks of 125,000 MC steps, while 5 blocks of 15,000-45,000 steps were
used for higher temperatures. 
The large hysteresis in the low temperature region indicates a
high number of metastable states, which is a characteristic of spin glasses.
These metastable states give rise to large error bars in
${\langle m^2 \rangle - \langle |m| \rangle^2}$  at low
temperatures, which are further accentuated by a division by $T$ in Eq.(4).
Although we feel that it might be possible to reduce the size of these
error bars, this would require very long runs and one must also ensure that
the system does not become trapped in one of these metastable states.
Nevertheless, despite the large error bars, a cusp-like feature in $\chi$
is clearly visible at $T_f\sim0.15$ for both the Penrose and
honeycomb lattices. We also find a saturation in this cusp
with respect to system size, which is consistent with spin glass behavior. 

In the insets of Fig.\ \ref{x.comp}(a) and (b) we have shown the quantity
$T \chi$, which
approaches a constant at high temperatures. Thus the high temperature phase is
strictly paramagnetic with $\chi$ obeying the Curie law.
In Fig.\ \ref{x.comp} we have shown the susceptibility function for a
quasiperiodic (Penrose) and a periodic (honeycomb) lattice. The general
temperature dependence of $T \chi$ for  the other two lattices, octagonal and
bathroom tile, is similar to that shown in Fig.\ \ref{x.comp}. 
Note that since the magnetic moment $\langle m \rangle \sim 0$ for all the
lattices, the quantity plotted in the insets of Fig.\ \ref{x.comp}(a) and (b)
is proportional to $\langle m^2 \rangle$. For a KT transition this quantity 
diverges below the transition temperature. For an Ising transition it remains
finite as the temperature approaches zero, and also is dependent on the system
size. Thus the fact that $\langle m^2 \rangle$ approaches zero for all the
lattices studied is an indication that they  do not exhibit either KT or
Ising-type transition as the temperature is lowered from the high temperature
paramagnetic phase. The saturation of the susceptibility with the system
size, the appearance of a kink-like feature and the vanishing of the 
function $\langle m^2 \rangle$ as the temperature approaches zero are
all consistent with the scenario of a low temperature spin glass phase.
As mentioned earlier the variation of $\langle m^2 \rangle$ or $T \chi$ with
temperature for the other two lattices, octagonal and bathroom tile,
is similar to that shown in the insets of Fig.\ \ref{x.comp}, although some
irregular features seem to appear in the function $\chi$ itself 
as the division by $T$, especially at low temperatures, strongly magnifies
minor deviations in $\langle m^2 \rangle$ from their exact values. For all
the four lattices studied $\langle m^2 \rangle$ rises, more or less smoothly,
from zero and saturates at a constant value as the temperature is increased.
Note that the bump in the inset of Fig.\ \ref{x.comp}(b) is probably due to a
convergence problem, as indicated by the large error bars.

It should also be noted that we have repeated the susceptibility calculation 
for the Penrose lattice without the frustration term $A_{ij}$ in Eq.(1).
We obtain a divergence in susceptiblity as the temperature is lowered,
consistent with a KT transition. By fitting the susceptibility to the form
proposed by Kosterlitz \cite {koster.74}
we have obtained a KT transition temperature $T_{KT}$ of  $1.027{\pm 0.002J}$
for this lattice and an exponent
in agreement with that obtained by Tobochnik and Chester \cite {toboch}
for the square lattice. 
The value of $1.027$ for $T_{KT}$ obtained by us for the Penrose lattice is
slightly higher than the value $(0.89-0.95)J$ reported in the literature
for the square lattice (see Tobochnik and Chester \cite {toboch} and 
references therein). Details of the unfrustrated Penrose lattice calculation
will be published elsewhere.

\subsection{Specific Heat}
Specific heats per site  for the two periodic and the two quasiperiodic
lattices obtained from the fluctuations in the energy U of the system:
\begin{equation}
\label{fluc.dis}
C=\frac{\left\langle U^2\right\rangle - \left\langle U\right\rangle^2}
{Nk_BT^2}\;,
\end{equation}
are  shown in Fig.\ \ref{cv.comp}. 
Results for various system sizes, clearly indicating a saturation with respect
to system size, are shown.
These results are averages between heating and cooling,
with the low temperature results having a somewhat larger hysteresis.
All averages were obtained after equilibriating, however, the high temperature
values were obtained by averaging over 5 blocks of 15,000 steps, whereas 5
blocks of 45,000 steps were used for the low temperatures.
The result for the unfrustrated Penrose lattice (without the $A_{ij}$ term in
Eq.(1)) is also shown (the inset of the Penrose section of Fig.\ \ref{cv.comp}).
Like the frustrated
case, the unfrustrated model shows a  saturation in specific heat with respect
to system size.
In the unfrustrated case the peak in the specific heat is at 1.10, which is
beyond the Kosterlitz-Thouless (KT) transition temperature $T_{KT}\sim 1.027$
obtained by us from the analysis of the susceptibility results.
The peaking of the specific heat at a temperature beyond the 
temperature at which a divergence appears in the linear susceptibility
is the expected behavior for a KT transition.  The specific heat
peak for the frustrated case is more rounded relative to the unfrustrated
case and occurs at a temperature lower than $T_{KT}$. This temperature is,
however, higher than the temperature $T_f$ at which, we believe,
the Edwards-Anderson
order parameter goes to zero or a cusp appears in the linear susceptibility.
The saturation in the peak height of the specific heat is a consequence of the
fact that it appears at a temperature at which the spin glass correlation is
finite.
Note that in both the unfrustrated and the frustrated cases the zero
temperature specific heat approaches a value of 0.5k$_B$.
This is consistent with 
the equipartition theorem valid for the Hamiltonian (1) with the cosine
function being truncated at the quadratic term.

A few comments regarding our results on the honeycomb lattice are in order at
this stage. 
Shih and Stroud \cite {shih}  carried out a Monte Carlo study of the
fully frustrated XY model on  the honeycomb lattice and reported the nature of
the transition as KT. Their study on the honeycomb lattice was carried out
together with the triangular lattice, and the conclusions regarding the nature
of the transition, Ising vs. KT, were primarily based on the saturation of the
specific heat with respect to system size. For the honeycomb lattice a         
saturation in the specific heat was obtained, while the triangular lattice did
not show any saturation. Consequently, the transitions were classified as KT and
Ising-type for the fully frustrated XY model on honeycomb and triangular  
lattices, respectively. Our 
work on the honeycomb lattice shows that the conclusion drawn by Shih and
Stroud\cite{shih} was premature, since the susceptibility for the
honeycomb lattice  shows no divergence
characteristic of a KT transition. Our results for the specific heat agree
numerically with the results of Shih and Stroud. In addition our results for
the system energy at the lowest temperature studied ($0.02J$) agree with
the ground state energy reported by Shih and Stroud \cite {shih}.
In TABLE I we present the average energy per spin for all the lattices and
cluster    
sizes used in the simulation. We find that with periodic boundary coditions
magnetic moments or ferromagnetic correlations decrease with increasing 
system size at low temperatures, yielding higher energy per spin for larger
sizes. For the 256 site honeycomb lattice our value of average energy per
site at $T=0.02J$, $-1.2169J$, is lower than the 
ground state energy $-1.2071J$ reported by Shih and Stroud \cite {shih}.
Presumably, Shih and Stroud report the value obtained for their
largest cluster of 576 sites. For a 1600 site cluster our value of the
energy at $0.02J$, $-1.198J$, is slightly higher.
In addition to specific heat and energy, our results for $q_{EA}$ are in good
agreement with the local order parameter values  reported by Shih and
Stroud \cite {shih}. 
Note  that the local order parameter studied by Shih and Stroud is the
square root of the order parameter $q_{EA}$ studied in this work.  
Based on these comparisons it appears that the study by Shih and Stroud
was correct in terms
of the accuracy of the quantities reported, but incomplete with regard to
correctly identifying the nature of the transition.

\subsection{Spin Glass Susceptibility}
In a ferromagnet, the approach to the ferromagnetic phase from temperatures 
above the Curie temperature $T_C$ is accompanied by a dramatic increase in
the range of
the spin correlations, which then diverges at $T_C$. A corresponding   
phenomenon occurs in spin glasses. However, it is not the spin correlation
function $\langle \vec S_i\cdot\vec S_j \rangle$, but rather its square that
acquires a long-range. This leads to the divergence, at the spin glass    
transition temperature $T_f$, of the spin glass susceptibility
\begin{equation}
\chi_{_{SG}} = \frac{1}{N} \sum_{ij}\langle \vec S_i\cdot\vec S_j \rangle^2
\hskip 0.8pc\relax (T > T_f)\:.
\end{equation}
$\chi_{_{SG}}$ satisfies a finite-size scaling relation of the
form \cite {bhatt}
\begin{equation}
\chi_{_{SG}} = L^{2-\eta}\bar\chi(L^{1/\nu}(T-T_f))\:,
\label{xsg.scale}
\end{equation}
where $\bar\chi$ is the scaling function, $L$ is the system length, $\nu$ is
the exponent for spin glass correlation length $\xi$ for $T \geq T_f$, and 
$\eta$ describes the power law decay of the spin glass correlation at $T_f$.
For the quasiperiodic lattices the linear dimensions in the $x$ and $y$
directions, $L_x$ and $L_y$, are different. Thus the scaling relations could
be studied using either of these two as a measure of the linear dimension of
the cluster. Alternatively, one could use $L=\sqrt{N}$ as a measure of the the
linear dimension of a cluster of N sites. All three options give similar
results for all the lattices we have
studied. In the following, we will present results with $L_x=L$ in Eq. (7).
To ensure a proper convergence of $\chi_{_{SG}}$, calculated via Eq.(6), we have
averaged over 5 blocks of 40,000-60,000 steps at low temperatures $(<0.2J)$ and
5 blocks of 60,000-80,000 steps at higher temperatures. By examining our
results every 5,000 steps, we find, for all lattices, little change in
$\chi_{_{SG}}$ over the
last 5,000-10,000 steps. Thus, we estimate that these chain lengths produce
at least a 95\% convergence in $\chi_{SG}$.

In Figs. (5-8) we show the scaling behavior of $\chi_{_{SG}}$ in
terms of $L_x$. In order to obtain estimates of $T_f\,$, we have been guided
by the exponents $\nu$ and $\eta$ reported by Bhatt and Young \cite {bhatt} in
their study of the $\pm J$
Ising  model, and by Jain and Young \cite {jain} for the $\pm J$ XY model, 
on square lattices.
We have also examined the Edwards-Anderson parameter in the
vicinity of $T_f$, where the relation $q_{EA}\sim (T-T_f)^\beta$ is supposed
to hold. Our overall finding is that the $\pm J$
Ising  model exponents $\nu$=2.6 $\pm 0.4$ and $\eta$=0.2 $\pm 0.05$ reported
by Bhatt and Young \cite {bhatt} provide 
a better fit to our data than the $\pm J$ XY model exponents
$\nu$=1.08 $\pm 0.27$ and $\eta$=0.3 $\pm 0.3$ given by Jain and Young 
\cite {jain}. Our results are summarized in TABLE II.
Below we examine our results case by case.

For the Penrose lattice, the Ising exponents (Fig.5(a)) fit the $\chi_{SG}$
data very well for $T_f=0.137J$ ( henceforth given in units of $J$). In
addition the slope of the log-log plot of $q_{EA}$ vs. $(T-T_f)$
(inset of Fig.5(a) and TABLE II) yields a value  $\beta=0.26$ in perfect
agreement with the hyperscaling relation $\beta=\eta\nu/2$. The XY exponents
$\eta=0.3$ and $\nu=1.08 $ also provide a reasonably good fit to the data
(for $T_f=0.11$), but the agreement between the value
of $\beta=0.19$ obtained from the simulation and the hyperscaling
value $\eta\nu/2=0.162$ is poorer. In Fig.5(b) we have chosen $\eta=0.26$,
and $\nu=1.35$, values within the error bars given for the XY model.
With $T_f=0.11$, the fit to the $\chi_{SG}$ data
is as good as for $\eta=0.3$ and $\nu=1.08 $, but these exponents 
($\eta=0.26$, $\nu=1.35$) yield $\eta\nu/2=0.1755$, in better agreement with
the value $\beta=0.19 \pm 0.01$ obtained from the order parameter results
(inset of Fig.5(b) and TABLE II).

For the octagonal lattice, the choice of $T_f=0.33$, $\nu=2.6$, and $\eta=0.3$
(Fig.6(a)) gives a good fit to the $\chi_{SG}$ data. Here the exponent $\eta$
is a little higher 
than the upper limit $0.25$, but $\nu$ is exactly the same as reported for the
Ising model. The order parameter yields $\beta=0.38$ (inset of Fig. 6(a) and
TABLE II), close to the hyperscaling value $0.39$. No reasonably good fit to
the $\chi_{SG}$ data can be obtained with $\nu$ and $\eta$ chosen around the
values given for the XY model. In Fig.6(b) we show the 
fit obtained with the XY exponents, $\eta=0.3$, and $\nu=1.35$. The best
results, obtained for $T_f=0.3$, are still quite poor in terms of satisfying
the scaling realtion. The order parameter plot (inset of Fig.6(b) and TABLE II)
yields $\beta=0.25$, while the hyperscaling value of $\beta$ is $0.2025$.

For the two periodic lattices, honeycomb and bathroom tile, the results
(Figs.7 and 8) are as follows.
With $\nu$ close to the Ising model value $2.6$ the best fit (Figs.7(a) and
8(a)) is obtained by choosing $\eta$ at the lower limit $0.15$ reported by
Bhatt and Young \cite {bhatt}.
The corresponding transition temperature $T_f$ in both cases is $0.12$. 
The log-log plots of the order parameter vs. $T-T_f$ yield $\beta=0.35$ and
$\beta=0.33$ (insets of Figs. 7(a) and 8(a)), respectively, for the honeycomb
and bathroom tile lattices. These values are about 40\% off the
corresponding hyperscaling values.   
With $\eta=0.29$ and $\nu=1.30$, chosen within the permitted values for the
XY model, the best fits for  both the periodic lattices are obtained for
$T_f=0.13$ (Figs. 7(b) and 8(b)). However, the order parameter data yield
$\beta=0.71$ and $\beta=0.50$ (insets of Figs. 7(b) and 8(b), and TABLE II)
for the honeycomb and the bathroom tile lattices, respectively. These values
are 3-4 times larger than the corresponding hyperscaling values $\eta\nu/2$.
Thus considering the $\chi_{SG}$
and the order parameter data together the agreement with the Ising model seems
better.

We have further examined the possibility of our spin glass susceptibility
data being consistent with $T_f$=0. 
Here we present results for the Penrose lattice only, the results for the
other lattices being qualitatively similar.
In Fig.9 we show the scaling plots of $\chi_{SG}$ with $T_f$=0 for the
$\pm J$ Ising and XY exponents, and also for the exponents given by our
best fit. Figs.9(a) and (b) show that there is a clear breakdown of the
scaling relation for $T_f$=0
if the exponents are close to those belonging to 
either of the $\pm J$ Ising or XY model. Fig.9(c) shows that  
substantial deviations in the values of $\eta$ and $\nu$ from the above two
models are needed to fit the $\chi_{SG}$ data with $T_f$=0. The values
$\nu$=9.6 and $\eta$=0.01 are far from the values reported  in the literature
for any spin glss model in 2D (see reference in Bhatt and Young \cite {bhatt}). 
Even if we accepted the values $\nu$=9.6 and $\eta$=0.01, our data for
susceptibility and Edwards-Anderson order parameter would be in conflict with
a zero transition temperature.
Thus based on an overall analysis of all our data we conclude that the
frustrated XY model on the 2D lattices considered exhibits spin glass
transitions at temperatures above zero.
It is not clear to us whether the  seemingly good agreement with the $\pm J$
Ising model exponents is purely coincidental, or there is a connection
with the results of Teitel and Jayaprakash \cite {teitel} on square lattices,
where the nature of the transition is found to change from KT to Ising type
as a result of including the frustration (the $A_{ij}$ term in Eq.(1)).

\section{Comparison with results on other lattices}
\label{sec:comp}
As stated in the introduction, the frustrated XY model has been studied on a
variety of  2D lattices, the most widely studied ones being the square and the
triangular. Although there is some controversy regarding the nature of the
transition on these
lattices, including the issue of the existence of more than one transitions,
most authors report the nature of the transition on these two lattices as being
'Ising-like'.
It will be useful to identify some feature of the model on the lattices
studied in the present work that distinguishes these from the square or
triangular lattice.
The obvious quantity to look at is the distribution of the lattice-dependent
vector potentials $A_{ij}$. Equivalently, we could rewrite Eq. (1) as 
\begin{equation}
H = \sum_{[ij]}-J\cos(A_{ij})\cos(\theta_{i}-\theta_{j})+\sum_{[ij]}
J\sin(A_{ij})\sin(\theta_{i}-\theta_{j}),
\end{equation}
where the first term is simply the standard XY Hamiltonian, with the 
lattice-dependent frustration appearing via the variation in the effective
nearest neighbor 
exchange parameters $\cos(A_{ij})$. We thus look at the distribution of the
effective coupling parameters, or simply the
quantity $\cos(A_{ij})$ for various lattices.

In Fig.\ \ref{Jij.comp} we show this quantity for 
the lattices studied in this work and 
in Fig.11 we show the same for the fully frustrated square and the triangular
lattices. The square and the triangular lattices, which show the Ising-like
transition, have much fewer values of the parameter $\cos(A_{ij})$ than all
the other lattices exhibiting spin glass behavior.
It should be noted that in order to compare the distribution of the quantity
$\cos(A_{ij})$ for various lattices, we chose, in each case, the reference   
$x$ and $y$ axes along some symmetry direction of the lattice. An arbitrary
choice of the reference axes, without any regard to the symmetry of the 
lattice, may result in a distribution showing spuriously large values of
the parameter $\cos(A_{ij})$. Such values of $\cos(A_{ij})$ give rise to
frustration which can be simply gauged away by rotating the reference frame
used to describe the lattice sites, and cannot be responsible for spin glass
behavior \cite {fradkin,toulouse}. 
In Fig.11 we also show the distribution of $\cos(A_{ij})$ for
a square lattice with an irrational flux, 
$f=\left(3-\sqrt{5}\right)/2$, through the plaquettes. This model was
studied by Halsey
\cite {halsey}, and was reported to show a low temperature spin glass phase.
We note that the distribution of $\cos(A_{ij})$ for this case is similar to
the distribution for the four cases studied by us, showing a large number of
possible values.

From the above discussion it appears that frustrated XY models with a wide
and more or less uniform distribution of the effective coupling parameters
(more appropriately with  a large number of values of the parameter
$\cos(A_{ij})$
distributed over the interval between -1 and 1) belong to a different
universality class than those with only a few possible values of $\cos(A_{ij})$.That the lattice structure is a relevant variable for the frustrated XY model
has been known for a long time. Here we have attempted to identify a common 
feature, dependent on the lattice structure as well as the magnetic field,
that might account for the       
spin/gauge glass phase in the frustrated XY model.  
 
It is of interest to compare our results with some work on 3D models.
Huse and Seung \cite {huse} have studied the so-called gauge or vortex glass
model on simple cubic lattices. The model is the same as that given by Eq.(1)
with the parameters $A_{ij}$ varying randomly and uniformly between $0$ and
$2\pi$.
These authors find a spin glass behavior with exponents that agree with the
$\pm J$ Ising spin glass exponents rather than the ones for the $\pm J$ 3D XY
model. These results are remarkably similar to ours.

\section{Comments and conclusions}
\label{sec:con}
In summary, we have shown, via the Edwards-Anderson order parameter, spin glass
susceptibility and linear susceptibility, the existence of a 
low temperature spin/guage  glass phase for the frustrated XY model
on two quasiperiodic (Penrose and octagonal) and two periodic (honeycomb and
bathroom tile) lattices. Our    
results for magnetization and specific heat also support this 
picture. We have also carried out a detailed study of the unfrustrated
ferromagnetic XY model on the Penrose lattice. The results are  
similar to that of a square lattice \cite {toboch}, with
a slightly higher KT transition temperature.

Our results for the quasiperiodic lattices are consistent with those
of Halsey \cite {halsey}, who finds a spin glass phase for the frustrated XY
model on a square lattice with an irrational flux through the plaquettes. Note
that for the Penrose, octagonal and the bathroom tile lattices we
can fully frustrate only one of the two elementary plaquettes at one time, 
the corresponding flux through the other plaquette being irrational. 
However, with the example of the honeycomb lattice we have shown that the
irrationality of the flux through the plaquettes is not a necessary condition
for the existence of the spin/gauge glass phase. It may, however, be a
sufficient 
condition. The common feature of all the cases studied is that the lattice
structure and the transverse magnetic field induce a large number of possible
values of the
effective coupling parameters $J\cos(A_{ij})$.               

The experimental implication of our study is that an array  
of Josephson junctions, forming any of the lattice structures discussed in this
work, in a suitably chosen transverse magnetic field  
should behave as a superconducting glass at low temperatures.
Advanced microfabrication techniques \cite {prober} should be capable of
generating such periodic/quasiperiodic arrays of superconducting grains. 
Experimental work of this kind has been reported \cite {gordon} on 2D fractal 
(Sierpinski-gasket) networks. 
Halsey \cite {halsey} has pointed out that for superconducting arrays with low
normal-state
resistivities the glass transition should basically appear as a mean-field
transition, with fluctuation effects being barely observable. For arrays
with high normal-state resistivities the fluctuation effects will cause
the glass transition to deviate substantially from a mean-field transition,  
with noticeable  system-dependent details.
As discussed by Ebner and Stroud \cite {ebner},
an important property of such glassy superconductors is a large difference
between their dc and ac susceptibilities.
 
It is interesting to note that  some 3D  quasicrystals
(both icosahedral and decagonal) are known to exhibit spin
glass phase \cite {berger,machado}. 
Of particular relevance to us are the decagonal quasicrystals, periodic in the
$z$-direction and quasiperiodic in the $xy$-plane.
A variation of the model studied in the present work may aptly describe
the magnetic properties of  some of these decagonal quasicrystals.

\acknowledgements
Financial support for this work was provided by  the Natural Sciences and 
Engineering Research Council of Canada. The authors are thankful to Prof. T.
Soma for providing them with the program to generate octagonal lattices.
A part of this work, containing some results for the Penrose and octagonal
 lattices, was presented at the Conference on Collective Phenomena
in Physics(CPiP'96)-Pattern Formation in Fluids and Materials, University
of Western Ontario, London, Ontario, CANADA (June 13-15, 1996) and is
scheduled to appear in a future issue of Physica A as Proceedings.

\begin{figure}
\caption{
Lattices used in the present study, along with their reference frames.
}
\label{lat.comp}
\end{figure}

\begin{figure}
\caption{
Edwards-Anderson order parameter as a function of temperature.
$N$ denotes the number of lattice sites and
$T_f$ is the estimate of the transition temperature for the corresponding
lattice, obtained from the analysis of spin glass susceptibility. }
\label{q.comp}
\end{figure}

\begin{figure}
\caption{
Linear susceptibility as a function of temperature
for the Penrose and the honeycomb lattices. The inset shows the product
$T\chi$, clearly indicating that the high temperature phase is paramagnetic,
with a Curie-like behavior for $\chi$.}
\label{x.comp}
\end{figure}

\begin{figure}
\caption{
Specific heat as a function of temperature. 
The inset in the Penrose tile shows the specific heat for the unfrustrated
Penrose lattice reaching its peak value slightly beyond the KT transition
temperature.}
\label{cv.comp}
\end{figure}

\begin{figure}
\caption{
Scaling plot of spin glass susceptibility for the Penrose
lattice using the exponents
(a) $\eta$=0.2 and $\nu$=2.6 reported by Bhatt and Young \protect{\cite{bhatt}}
in their study of the $\pm J$ Ising model;
(b) $\eta$=0.26 and $\nu$=1.35, chosen within the intervals $\nu$= 1.08$\pm$
0.27 and $\eta$=0.3$\pm$0.3
reported by Jain and Young \protect{\cite{jain}} in their study of $\pm J$ XY
model. The insets show the log-log plot of the Edwards-Anderson order
parameter versus $(T_f-T)$ with $T_f=0.137$ for case (a) and $T_f=0.11$ for
(b), obtained from the fit to the scaling relation,
Eq.~\protect{\ref{xsg.scale}}.  
The slope of this plot in (a) yields a value of 0.26
for the exponent $\beta$, in  perfect agreement with the hyperscaling
relation $\beta =\nu\eta/2$. In (b) the slope is 0.19, 
close to the value 0.1755 given by the hyperscaling relation.
See text and TABLE II for details.}
\label{pen.chisg}
\end{figure}

\begin{figure}
\caption{
Scaling plot of spin glass susceptibility for the octagonal 
lattice using the exponents (a) $\eta$=0.3 and $\nu$=2.6; and
(b) $\eta$=0.3 and $\nu$=1.35. The slope of the log-log plot in the inset of
(a) is 0.38, in agreement with the hyperscaling relation
$\beta =\nu\eta/2$.   
The slope of the log-log plot in the inset in (b) is 0.25, whereas the
hyperscaling value is 0.2025. See text and TABLE II for details. }
\label{oct.chisg}
\end{figure}

\begin{figure}
\caption{
Scaling  plot of  spin glass susceptibility for the honeycomb lattice.
See text and TABLE II for details with regard to agreement with $\pm J$ Ising
and XY spin glass models.}
\label{hon.chisg}
\end{figure}

\begin{figure}
\caption{
Scaling plot of spin glass susceptibility for the bathroom tile lattice.
See text and TABLE II for details with regard to agreement with $\pm J$ Ising
and XY spin glass models.}
\label{bat.chisg}
\end{figure}

\begin{figure}
\caption{
Scaling  plot of  spin glass susceptibility for the Penrose lattice
with $T_f$=0,
using (a) $\pm J$ Ising, (b) $\pm J$ XY exponents, and (c) exponents
obtained from the best fit to Eq.~\protect{\ref{xsg.scale}}.}
\label{pen.tf0}  
\end{figure}

\begin{figure}
\caption{ 
Distribution of the effective nearest neighbor coupling parameter $\cos(A_{ij})$
for the lattices used in this work.}
\label{Jij.comp}
\end{figure}

\begin{figure}
\caption{   
Distribution of the effective nearest neighbor coupling parameter
$\cos(A_{ij})$ for the fully frustrated triangular and square lattices,
and for square lattice
with an irrational flux, $f=\left(3-\protect{\sqrt{5}}\right)/2$, through the
plaquettes.}
\end{figure}

\newpage

\begin{table}
\caption{Average energy per spin(site)
$\langle u\rangle$  at the lowest temperature, $0.02J$,
used in the simulation for various lattices and cluster sizes, $N$. $L_x$
and $L_y$ denote the linear dimensions of the clusters in the $x$ and
$y$ directions, respectively.
\label{table1}}
\begin{tabular}{lllll}
Lattice & $N$ & $L_x$ & $L_y$ & $\langle u\rangle$\\
\tableline
Penrose & 644       &  24.80 & 21.09 & -1.4938\\
        & $1\,686$  &  40.12 & 34.13 & -1.4893\\
        & $4\,414$  &  64.92 & 55.23 & -1.4889\\
        & $11\,556$ & 105.05 & 89.36 & -1.4884\\
\tableline
Octagonal & 239      & 14.07 & 14.07 & -1.3878\\
          & $1\,393$ & 33.97 & 33.97 & -1.3926\\
          & $8\,119$ & 82.01 & 82.01 & -1.3947\\
\tableline
Honeycomb & 242      &  19.05 &  19.05 & -1.2169\\
          & $1\,682$ &  50.23 &  50.23 & -1.1980\\
          & $4\,050$ &  77.94 &  77.94 & -1.1926\\
          & $8\,192$ & 110.85 & 110.85 & -1.1868\\
\tableline
Bathroom Tile & 256      &  19.31 &  19.31 & -1.1842\\
              & $1\,600$ &  48.28 &  48.28 & -1.1846\\
              & $4\,096$ &  77.25 &  77.25 & -1.1843\\
              & $8\,100$ & 108.64 & 108.64 & -1.1847\\
\end{tabular}
\end{table}

\begin{table}
\caption{
Spin glass transition temperature $T_f$ and the exponents $\eta$  and $\nu$
obtained for the various lattices from  finite size scaling analysis of
the spin glass susceptibilty. The exponent $\beta$, obtained from the log-log
plot of the Edwards Anderson order parameter $q_{EA}$ and $T-T_f$ is also shown.
The division into the categories "Ising" and "XY" is based on the proximity of
the exponents $\eta$ and $\nu$ to the values reported by Bhatt and Young
\protect{\cite{bhatt}} and Jain and Young \protect{\cite{jain}} in their study
of $\pm J$ Ising and XY models, respectively, on square lattices.
\label{table2}}
\begin{tabular}{l|llll|llll}
& \multicolumn{4}{c}{Ising\tablenotemark[1]} & 
\multicolumn{4}{c}{XY\tablenotemark[2]}\\
Lattice & $T_f$ & $\eta$ & $\nu$ & $\beta$ & $T_f$ & $\eta$ & $\nu$ & $\beta$\\
\tableline
Penrose & 0.137 & 0.20 & 2.6 & 0.26 & 0.11 & 0.26 & 1.35 & 0.19\\
Octagonal & 0.33 & 0.3 & 2.6 & 0.38 &0.3 &0.3 &1.35 &0.25 \\
Honeycomb & 0.12 & 0.15 & 2.8 & 0.35 & 0.13 & 0.29 & 1.30 & 0.71\\
Bathroom Tile & 0.12 & 0.15 & 2.6 & 0.33 & 0.13 & 0.29 & 1.30 & 0.50 \\ 
\end{tabular}
\tablenotetext[1]{$\pm J$ Ising Exponents in Bhatt and Young$^6$ : $\eta=0.2 \pm 0.05$, $\nu=2.6 \pm 0.4$, $\beta=\eta\nu/2=0.26$. }
\tablenotetext[2]{ $\pm J$ XY Exponents in Jain and Young$^7$  are  $\eta=0.3 \pm 0.3$, $\nu=1.08 \pm 0.27$, $\beta=\eta\nu/2=0.162$.}
\end{table}
\end{document}